\documentclass[aip,rsi,amsmath,amssymb,reprint]{revtex4-1}

\usepackage{graphicx,caption,subcaption}
\usepackage{hyperref}
\usepackage[capitalise]{cleveref}
\usepackage[normalem]{ulem} 
\usepackage{xcolor} 
\usepackage{multirow,booktabs}
 
\begin{document}

\title{Anharmonic infrared spectra of cationic pyrene and superhydrogenated derivatives} 

\author{Zeyuan Tang}
\affiliation{School of Chemistry and Chemical Engineering, Hainan University, Haikou, 570228, China}
\affiliation{Center for Interstellar Catalysis (InterCat), Department of Physics and Astronomy, Aarhus University, Ny Munkegade 120, 8000 Aarhus C, Denmark}

\author{Frederik G. Doktor}
\affiliation{Center for Interstellar Catalysis (InterCat), Department of Physics and Astronomy, Aarhus University, Ny Munkegade 120, 8000 Aarhus C, Denmark}

\author{Rijutha Jaganathan}
\affiliation{Center for Interstellar Catalysis (InterCat), Department of Physics and Astronomy, Aarhus University, Ny Munkegade 120, 8000 Aarhus C, Denmark}

\author{Julianna Palot{\'a}s}
\affiliation{Radboud University, Institute for Molecules and Materials, FELIX Laboratory, Toernooiveld 7,
6525ED Nijmegen, The Netherlands}
\affiliation{School of Chemistry, University of Edinburgh, Joseph Black Building, Kings Buildings, David Brewster Road, Edinburgh EH9 3FJ, United Kingdom}

\author{Jos Oomens}
\affiliation{Radboud University, Institute for Molecules and Materials, FELIX Laboratory, Toernooiveld 7,
6525ED Nijmegen, The Netherlands}

\author{Liv Hornekær}
\affiliation{Center for Interstellar Catalysis (InterCat), Department of Physics and Astronomy, Aarhus University, Ny Munkegade 120, 8000 Aarhus C, Denmark}
\affiliation{Interdisciplinary Nanoscience Center (iNANO), Aarhus University, Gustav Wieds Vej 14, 8000 Aarhus C, Denmark}

\author{Bjørk Hammer}
\email{hammer@phys.au.dk}
\affiliation{Center for Interstellar Catalysis (InterCat), Department of Physics and Astronomy, Aarhus University, Ny Munkegade 120, 8000 Aarhus C, Denmark}

\date{\today}

\begin{abstract}
Studying the anharmonicity in the infrared (IR) spectra of polycyclic aromatic hydrocarbons (PAHs) at elevated temperatures is important to understand vibrational features and chemical properties of interstellar dust,
especially in the James Webb Space Telescope (JWST) era.
We take pyrene as an example PAH and investigate how different degrees of superhydrogenation affects the applicability of the harmonic approximation and the role of temperature in IR spectra of PAHs.
This is achieved by comparing theoretical IR spectra generated by classical molecular dynamics (MD) simulations and experimental IR spectra obtained via gas-phase action spectroscopy which utilizes the Infrared Multiple Photon Dissociation (IRMPD).
All simulations are accelerated by a machine learning interatomic potential, in order to reach first principle accuracies while keeping low computational costs.
We have found that the harmonic approximation with empirical scaling factors is able to reproduce experimental band profile of pristine and partially superhydrogenated pyrene cations.
However, a MD-based anharmonic treatment is mandatory in the case of fully superhydrogenated pyrene cation for matching theory and experiment.
In addition, band shifts and broadenings as the temperature increases are investigated in detail.
Those findings may aid in the interpretation of JWST observations on the variations in band positions and widths of interstellar dust.
\end{abstract}

\maketitle

\section{Introduction}
Polycyclic aromatic hydrocarbons (PAHs) are generally regarded as the carriers of infrared emission bands (so-called aromatic infrared bands, AIBs) at 3.3, 6.2, 7.7, 8.7, 11.3 and 12.7 $\mu$m, 
which are observed in a wide variety of astronomical objects \cite{tielens2008,peeters2021}.
The launch of the James Webb Space Telescope (JWST) allows studying the detailed band profiles of AIBs with unprecedented accuracy.
For example, a JWST study of AIBs in the Orion Bar by the PDR4ALL team has pointed out the importance of anharmonicity in the understanding of 5.25, 6.2 and 11.2 $\mu$m features \cite{chown2024}.
Another JWST program aims at anharmonic features (e.g. overtone and combination bands in PAHs) in the 1-5 $\mu$m region \cite{boersma2023}.
The study of anharmonicity in PAH spectra from laboratory astrophysics and quantum chemical calculations is needed in order to aid the interpretation of on-going and future JWST observations on AIBs.

Pyrene, the smallest compact PAH, has been extensively studied for decades in laboratory astrophysics \cite{joblin1995, chakraborty2019} and quantum chemical calculations \cite{mackie2016,chen2018a}, for a better understanding of anharmonicity.
Previous investigations of anharmonic effects include pyrene in neutral \cite{joblin1995,mackie2016}, cationic \cite{panchagnula2020} and anionic \cite{salzmann2024} states, each of which may contribute to different bands in AIBs.
The understanding of anharmonicity is often achieved by comparing experimental and theoretical (both harmonic and anharmonic) IR spectra.
One of the most common methods of computing anharmonic IR spectrum is the generalized vibrational perturbation theory to the second order (GVPT2) \cite{barone2005,SNSD}.
It is built on top of the harmonic approximation and normally truncates the expression of potential energy to fourth order.
Resonances between vibrational states, which are important for the 3.3 $\mu$m C-H stretching band \cite{maltseva2015,mackie2015}, could also be included in such a method.
GVPT2 with the density functional theory (DFT) accuracy has been widely used in computing anharmonic IR spectra of a wide range of PAHs \cite{mackie2015,mackie2016,mackie2018b,esposito2024a}, which is normally capable of reproducing most vibrational features in experiments.
The largest carbonaceous molecule that has been studied with GVPT2 at the DFT level so far is the protonated C$_\text{70}$ \cite{salzmann2024}.

Gas-phase PAHs like pyrene could reach around 1000 K after absorbing a single UV photon under certain interstellar conditions \cite{joblin1995}.
The role of temperature in the IR spectra of PAHs including pyrene has been studied experimentally both in the gas phase \cite{joblin1995} and in the solid phase \cite{chakraborty2019}.
A common observation is that bands shift to lower wavenumbers and get broader at higher temperatures.
Theoretically, a full quantum method based on GVPT2 and a dynamical approach via classical molecular dynamics (MD) have been tried to study the temperature-dependent IR spectrum of gas-phase pyrene \cite{chakraborty2021}.
The first method starts with a GVPT2 calculation to obtains third and fourth derivatives of potential energy surface at a local minimum configuration, and resonance states \cite{mulas2018,mackie2018a}.
Afterwards, the Wang-Landau method \cite{wang2001} is used to do statistical sampling of vibrational states, and estimates the spectrum of an equilibrium system in the microcanonical ensemble.
The obtained spectrum under a Laplace transform can be converted into an anharmonic vibrational spectrum at a given temperature in the canonical ensemble \cite{basire2009}.
This method becomes computationally expensive at high temperatures due to increased sampling space (e.g. the huge number of resonating states) \cite{chakraborty2021}.

The other method is built on MD simulations which involve time propagations of physical properties, such as positions, velocities and dipole moments.
No assumptions or truncations of PES are needed in this theoretical framework. 
IR spectra can be obtained from the Fourier transform of the dipole autocorrelation functions.
Therefore, the MD-based approach naturally includes anharmonic effects (e.g. combination bands and asymmetric broadenings), while temperature could be controlled by a thermostat.
A statistically converged spectrum often requires millions of MD steps, which are computationally expensive with accurate electronic structure methods (e.g. hybrid DFT for PAH spectroscopy).
One possible solution is to use faster but less accurate methods like density functional based tight binding (DFTB) \cite{chakraborty2021} and PM3 empirical methods \cite{chen2019a}.
They have been used to study the temperature evolution of band positions and widths in IR spectra of PAHs.
It is also feasible to run fast MD simulations while keeping the DFT accuracy, with the help of machine learning interatomic potential (MLIP) \cite{gastegger2017,behler2021}.
The combination of MLIP and MD, namely MLMD, has successfully been applied to compute anharmonic IR spectra of PAHs \cite{mai2025}.
MLMD has also been used to study the thermal evolution in IR spectra of nanosilicate clusters \cite{tang2023}, which are another type of interstellar dust other than PAHs.
In addition, the MLIP could be combined with the global structure optimization \cite{GOFEE,christiansen2022,ronne2022} to study energetically favorable structures of interstellar dust analogs \cite{tang2022a,slavensky2023,slavensky2024}. 

Some AIB ratios, such as the relative intensity between the 3.3 and 3.4 $\mu$m band, are sensitive to the composition of aromatic vs aliphatic carbon in PAHs.
Therefore, the IR spectroscopy of superhydrogenated PAHs (HPAHs) which normally have mixed aromatic and aliphatic carbons, has been a focus of many experimental and theoretical studies \cite{mackie2018b,maltseva2018,esposito2024}.
An accurate band assignment of HPAHs requires both high-resolution IR experiments and anharmonic IR calculations.
However, a systematic investigation on how the increased degree of superhydrogenation affects anharmonicity in the IR spectroscopy of HPAHs, is still limited.
This motivates our choices of the investigated PAHs which include pristine, partially superhydrogenated and fully superhydrogenated pyrene cations.
We will present a joint experimental and theoretical study on the anharmonicity in the IR spectroscopy of the investigated PAHs.
The paper is organized as follows.
We first introduce the experimental and theoretical approaches to obtain IR spectra of the investigated PAHs.
Experimental IR spectra are measured by the infrared multiple-photon dissociation (IRMPD) \cite{IRMPD_review,IRMPD_tutorial_review}, and are used to calibrate the following anharmonic calculations.
In the theoretical method part, we will emphasize the MD-based method to compute anharmonic IR spectra which are temperature dependent.
All MD simulations will be performed with a MLIP.
This section includes our choices of machine learning architecture and our sampling strategy of training data.
We move on to discuss the accuracy and the validation of the trained MLIP on three target properties (energy and atomic forces for driving MD simulations, dipole moments for IR postprocessing).
The spectroscopic performance of the trained MLIP will then be examined by its comparison with the DFT-calculated IR spectra.
Afterwards, we investigate the anharmonic effects in the IR spectra of superhydrogenated pyrene cations, through the comparison between the experimental spectra and the theoretical spectra (both within the harmonic approximation and from the MD-based approach).
The role of temperature will be studied via the variations of band positions and widths of computed IR spectra from MD simulations.
Finally, the astrophysical implications of anharmonicity will be discussed.

\section{Method}
\begin{figure}
	\centering
	\includegraphics[width=1\linewidth]{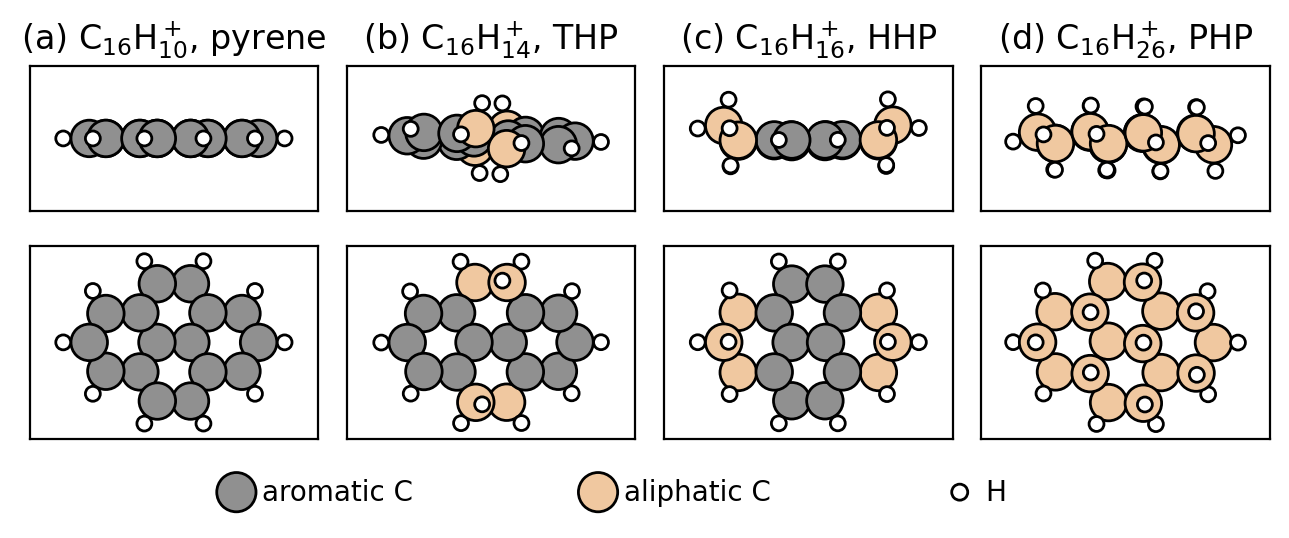}
	\caption{Top and side views of PAH cations: (a) pyrene (C$_{16}$H$_{10}^+$),  (b) 4,5,9,10-tetrahydropyrene (C$_{16}$H$_{14}^+$, THP), 
    (c) 1,2,3,6,7,8-hexahydropyrene (C$_{16}$H$_{16}^+$, HHP) and (d) perhydropyrene (C$_{16}$H$_{26}^+$, PHP).
	Color codes are: gray (aromatic carbon), light yellow (aliphatic carbon), white (hydrogen).
	}
	\label{fig:snaps-pyrenes}
\end{figure}

The PAH cations investigated in this work are pyrene (C$_{16}$H$_{10}^+$), 4,5,9,10-tetrahydropyrene (C$_{16}$H$_{14}^+$, THP), 
1,2,3,6,7,8-hexahydropyrene (C$_{16}$H$_{16}^+$, HHP) and perhydropyrene (C$_{16}$H$_{26}^+$, PHP), 
with their structure snapshots given in \ref{fig:snaps-pyrenes}.
The selection of four pyrene-based PAH cations allows studying how the degree of superhydrogenation (pristine, partially superhydrogenated and fully superhydrogenated) affect anharmonicity in IR spectra of PAHs.

\subsection{Experimental setup} 
To obtain IR spectra of the differently superhydrogenated PAH cations, experiments were carried out at the Free Electron Lasers for Infrared eXperiments (FELIX) facility, Radboud University, in Nijmegen, The Netherlands.
A tandem mass spectrometry (MS/MS) setup for gas-phase action spectroscopy utilizes a modified 3-D quadrupole ion trap mass spectrometer (Bruker Amazon Speed ETD) that allows IR radiation from FELIX to enter and interact with the trapped ions.
For a more detailed description of the setup see Martens et al. \cite{martens2016}

The setup employs Infrared multiple-photon dissociation (IRMPD) to fragment the trapped ions and spectroscopic information may be gained from the degree of fragmentation as a function of IR wavelength \cite{IRMPD_review,IRMPD_tutorial_review}.
The IRMPD process increases the internal energy of the molecule above the dissociation threshold, through intra-molecular vibrational redistribution (IVR), eventually leading to fragmentation via the lowest energy dissociation channel(s).
FELIX produces intense pulses of IR photons in a wavelength range between 3 and 150 $\mu$m.
The IR radiation is delivered in 5 - 10 $\mu$s macropulses at 10 Hz with an approximate macropulse energy of 40 - 100 mJ/pulse and a bandwidth of 0.4 \% of the set wavelength.
Each macropulse consists of a train of up to 6\,ps long micropulses spaced by 1 ns.
After irradiation of the ions at the desired wavelength, the IRMPD-induced fragment ions and remaining precursor ions are scanned out of the ion trap on the basis of their mass/charge ratio (\textit{m/z}) and detected.
The mass spectra recorded for each individual wavelength are then used to generate a surrogate IR vibrational spectrum over the wavelength range via the yield ratio of fragments over all ions:
\begin{equation}\label{eq:yield IR1}
\text{Yield($\lambda$)} = \frac{\sum I_{\text{fragments}}}{\sum I_{\text{all ions}}}. 
\end{equation}
From this, the fragment fluence, $S$, which is proportional to the absorption cross section at the given wavelength, is obtained according to \cref{eq:yield IR} \cite{berden2019}:
\begin{equation}\label{eq:yield IR}
S(\lambda) = -\text{ln}(1-\text{Yield})\\
\end{equation}
The IR spectra presented in this work are obtained by plotting $S$ as a function of wavenumber.

The HPAH cations examined experimentally in this study are 4,5,9,10-tetrahydropyrene (C$_{16}$H$_{14}$, THP) ($>$98\%; TCI), 1,2,3,6,7,8-hexahydropyrene (C$_{16}$H$_{16}$, HHP) ($>$98\%; TCI) and perhydropyrene (C$_{16}$H$_{26}$, PHP) ($>$97\%; TCI).
The HPAHs were dissolved in toluene creating a 1 mmol stock solution, and then diluted further in a 50\%/50\% methanol-toluene mixture.
Cations of the PAHs were formed in an atmospheric pressure chemical ionization (APCI) source which has been successfully used to study large carbonaceous species such as PAHs \cite{lien2007,palotas2021a} and fullerene derivatives \cite{palotas2020,palotas2021,palotas2022,finazzi2024}.

FELIX was set to produce IR radiation in the range of 600 - 1800 cm$^{-1}$ (5.5 $\mu$m - 16.6 $\mu$m) and frequency scans were performed with a step size of 3 cm$^{-1}$.
All IR spectra were averaged multiple times to increase the signal to noise ratio, followed by a linear power correction to account for the number of pulses and the varying power of FELIX across the wavelength range.
\subsection{Computational setup} 

\subsubsection{MLIP}
The MLIP in this work is represented by the polarizable atom interaction neutral network (PaiNN) \cite{PaiNN}, which has an equivariant message passing architecture.
The atomic representations are initialized to learnable embeddings with a feature dimension of 128, according to element types (carbon and hydrogen in this work).
They are then passed though 5 interaction blocks for sharing and updating information of local chemical environments.
A cutoff of 5 Å is used for determining neighbor lists and local environments.
Other architecture details of PaiNN are described elsewhere \cite{PaiNN}.
Since the motivation of developing such a PaiNN model is to run MD simulations and extracting IR spectra from dipole autocorrelation functions, 
we have utilized a single PaiNN model for predicting energy $\hat{\text{E}}$, atomic forces $\hat{\textbf{F}}_i$ and dipole moment $\hat{\bm{\mu}}$.
A combined loss function is defined as \cref{eq-loss}
\begin{align} \label{eq-loss}
	\begin{split}
	\mathcal{L}_{\text{E,F,}\mu} =& \sum_{n=1}^{N_\text{structures}} \lambda_\text{E} || \text{E}^\text{ref} - \hat{\text{E}} ||^2 \\
	+& \frac{\lambda_\text{F}}{3\text{N}_\text{atoms}} \sum_{i=1}^{\text{N}_\text{atoms}} || \textbf{F}^\text{ref}_i - \hat{\textbf{F}}_i ||^2 \\
	+& \frac{\lambda_\bm{\mu}}{3} || \bm{\mu}^\text{ref} - \hat{\bm{\mu}} ||^2
	\end{split}
\end{align}
where the trade-off between energy, atomic forces and dipole moment is set to $\lambda_\text{E} = 0.05$, $\lambda_\text{F} = 0.9$ and $\lambda_\bm{\mu} = 0.05$.

We have trained separate PaiNN models for pyrene, THP, HHP and PHP.
The training data are obtained by the normal mode sampling \cite{ANI-1,braams2009,qu2018} and the furthest point sampling \cite{bartok2017,imbalzano2018} as described in our previous work \cite{tang2023}.
The number of training data is 18279, 21091, 22497 and 29527 for pyrene, THP, HHP and PHP, respectively.
The training data are split into a training subsets and a validation subsets with a 9:1 ratio.

\subsubsection{DFT calculations}
All density functional theory calculations were performed with the B3LYP functional \cite{lee1988,becke1993a} and the def2-TZVP basis set \cite{def2}, as implemented in ORCA 5.0.3 \cite{ORCA4,ORCA5}.
B3LYP calculations were accelerated by the RIJCOSX approximation \cite{RIJCOSX}.
The dispersion correction was treated by an empirical dispersion scheme DFT-D3 \cite{D3} and the Becke-Johnson damping \cite{D3BJ}.
A tight SCF convergence criterial "TightSCF" was selected.

\subsubsection{MLMD simulations}
Molecular dynamic simulations were performed using the MLIP, and these simulations are referred as MLMD.
A time step of 0.5\,fs was used in all MLMD.
The temperature was controlled by the Nosé-Hoover chain thermostat \cite{martyna1992} with the time constant of 100 fs and a chain length of 3.
The total simulation time was set to 55\,ps, of which the first 5 ps was used for equilibration. 
The remaining 50 ps trajectory was used for extracting IR spectra via the Fourier transform of the autocorrelation function of the dipole time derivative \cite{thomas2013} according to \cref{eq:autocorr}.
\begin{equation}\label{eq:autocorr}
	I_{IR} \propto \int _ {-\infty} ^ {+\infty} \langle	\dot{\mu}(\tau) \dot{\mu}(\tau+t) \rangle_{\tau} e^{-i \omega t} dt
\end{equation}
Other IR postprocessing details can be found in Tang et al. \cite{tang2023}
In order to improve the statistics of the obtained IR spectra, 10 independent MLMD runs were performed for each system with different initialization of velocities.
Each IR spectrum in the following section was averaged over 10 MLMD runs.
All MLMD simulations and IR processing were done with SchNetPack v 2.0.4 \cite{SchNetPack,SchNetPack2}.

A 55\,ps MLMD simulation of the PHP cation (42 atoms) took only about 1 hour on a single NVIDIA V100 GPU card.
If the same simulation were performed with DFT, the estimated run time would be about 305 days on 32 CPU cores (Intel Xeon Gold 6240),
as the single point calculation of the PHP cation took around 4 minutes at the B3LYP/def2-TZVP D3(BJ) level.
The GPU training of the PHP cation only took around 3 days on a single NVIDIA V100 GPU card, which is negligible compared to the estimated run time of a DFT-MD simulation.
This shows a tremendous speedup of MLMD over DFT-MD for computing anharmonic IR spectra of large PAH cations.

\section{Results}

\subsection{MLIP benchmark}
The accuracy of the trained MLIP is assessed by computing mean absolute error (MAE) and root mean squared error (RMSE)
between MLIP predictions and DFT values. \cref{tab:painn-accuracy} reports the MAE and RMSE for three properties (energy, atomic forces and dipole moment)
in four systems (pyrene, THP, HHP and PHP).
The largest error across training and validation data sets is 0.3 meV/atom for energy, 51.8 meV/Å for atomic forces and 0.032 Debye for dipole moment,
showing the excellent accuracy of the trained MLIP.
\begin{table*}
	\centering
	\caption{Accuracy of the trained MLIP on training and validation data sets using two metrics, 
	namely mean absolute error (MAE) and root mean squared error (RMSE)}
	\label{tab:painn-accuracy}
	\begin{tabular}{lcccccccccccc}
		\toprule[0.05cm]
		\multirow{3}{*}{System}
		& \multicolumn{4}{c}{Energy [meV/atom]} & \multicolumn{4}{c}{Atomic forces [meV/Å]} & \multicolumn{4}{c}{Dipole monent [Debye]} \\ \cmidrule[0.03cm](l{5pt}r{5pt}){2-5} \cmidrule[0.03cm](l{5pt}r{5pt}){6-9} \cmidrule[0.03cm](l{5pt}r{5pt}){10-13} 
		& \multicolumn{2}{c}{Training} & \multicolumn{2}{c}{Validation} & \multicolumn{2}{c}{Training} & \multicolumn{2}{c}{Validation} & \multicolumn{2}{c}{Training} & \multicolumn{2}{c}{Validation} \\
		\cmidrule[0.03cm](l{5pt}r{5pt}){2-3} \cmidrule[0.03cm](l{5pt}r{5pt}){4-5} \cmidrule[0.03cm](l{5pt}r{5pt}){6-7} \cmidrule[0.03cm](l{5pt}r{5pt}){8-9} \cmidrule[0.03cm](l{5pt}r{5pt}){10-11} \cmidrule[0.03cm](l{5pt}r{5pt}){12-13} 
		& MAE & RMSE & MAE & RMSE & MAE & RMSE & MAE & RMSE & MAE & RMSE & MAE & RMSE \\ \cmidrule[0.03cm](l{5pt}r{5pt}){1-13}
	   pyrene & 0.1 & 0.1 & 0.1 & 0.1 & 5.5 & 7.6 & 7.7 & 16.2 & 0.005 & 0.007 & 0.007 & 0.010 \\
	   THP & 0.1 & 0.2 & 0.2 & 0.3 & 9.7 & 13.7 & 13.8 & 28.3 & 0.010 & 0.014 & 0.012 & 0.021 \\
	   HHP & 0.1 & 0.1 & 0.1 & 0.2 & 7.6 & 10.6 & 9.8 & 15.5 & 0.007 & 0.009 & 0.009 & 0.012 \\
	   PHP & 0.2 & 0.3 & 0.2 & 0.3 & 15.6 & 22.1 & 23.7 & 51.8 & 0.014 & 0.019 & 0.020 & 0.032 \\
	\bottomrule[0.05cm]
\end{tabular}
\end{table*}

\begin{figure}
	\begin{center}
	\includegraphics[width=\columnwidth]{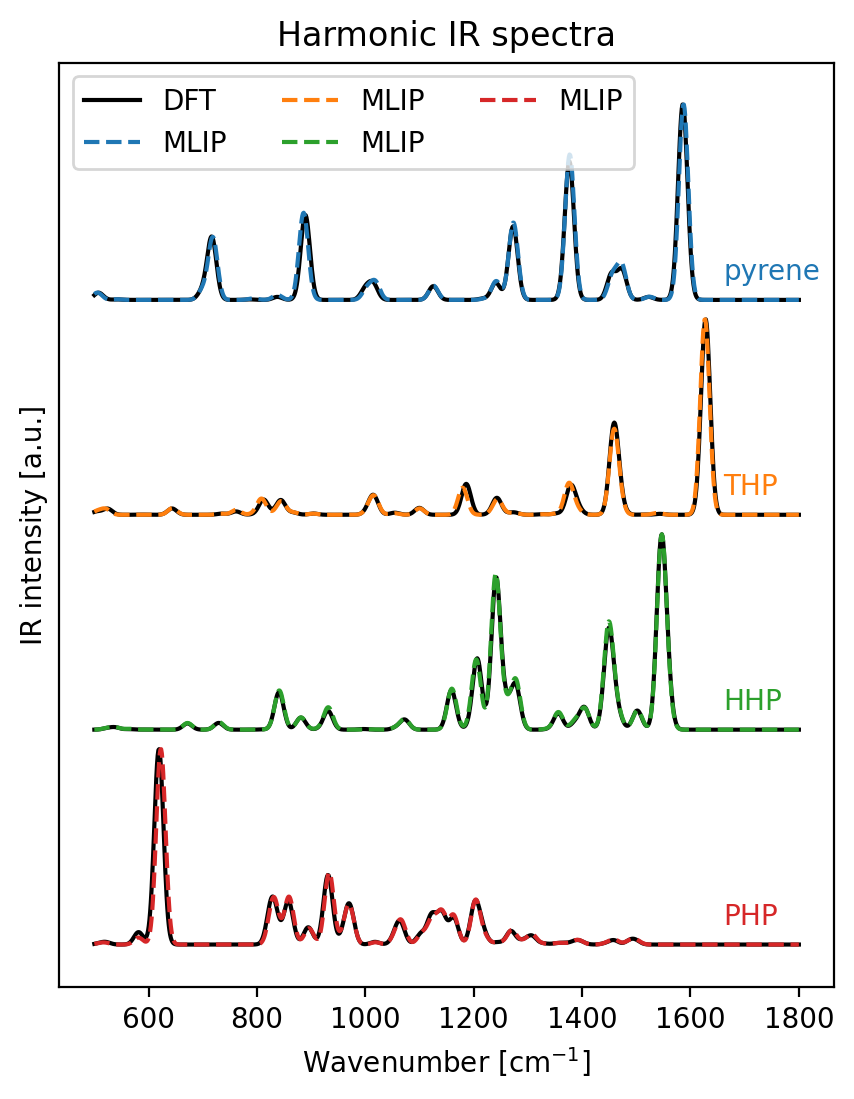}
	\caption{IR spectra of pyrene, THP, HHP and PHP radical cations within the harmonic approximation.
	Solid lines are DFT results while colored dashed lines are generated with MLIP.
	}
    \label{fig:harmonic-ir}
    \end{center}
\end{figure}

The spectroscopic behavior of the trained MLIP is examined by the calculation of harmonic IR spectra.
The predicted harmonic frequencies and IR intensities by the MLIP are compared with DFT-calculated results, as shown in \cref{fig:harmonic-ir}.
The excellent agreement between MLIP and DFT is achieved in the harmonic IR spectra of all four PAHs investigated in this work.
\cref{tab:painn-harmonic} gives quantitative error measurements for comparing MLIP and DFT results.
The highest MAE/RMSE in four PAHs is 3.9/2.7 cm$^{-1}$ for harmonic frequency and 12.0/5.0 km/mol for IR intensity.

\begin{table}
	\centering
	\caption{Spectroscopic behavior of the trained MLIP for computing IR spectra within the harmonic approximation}
	\label{tab:painn-harmonic}
	\begin{tabular}{ccccc}
		\toprule[0.05cm]
		\multirow{2}{*}{System}
		& \multicolumn{2}{c}{Harmonic frequency [cm-1]} 
		& \multicolumn{2}{c}{IR intensity [km/mol]} \\ 
		\cmidrule[0.03cm](l{5pt}r{5pt}){2-3} \cmidrule[0.03cm](l{5pt}r{5pt}){4-5}
		& MAE & RMSE & MAE & RMSE \\ \cmidrule[0.03cm](l{5pt}r{5pt}){1-5}
	pyrene & 2.7 & 3.9 & 1.3 & 3.6 \\
	THP & 1.7 & 2.4 & 3.3 & 12.0 \\
	HHP & 0.7 & 1.6 & 0.7 & 1.4 \\
	PHP & 1.1 & 1.4 & 5.0 & 10.6 \\
	\bottomrule[0.05cm]
	\end{tabular}
\end{table}

\subsection{Applicability of the harmonic approximation}
\begin{figure*}
	\begin{center}
	\includegraphics[width=\textwidth]{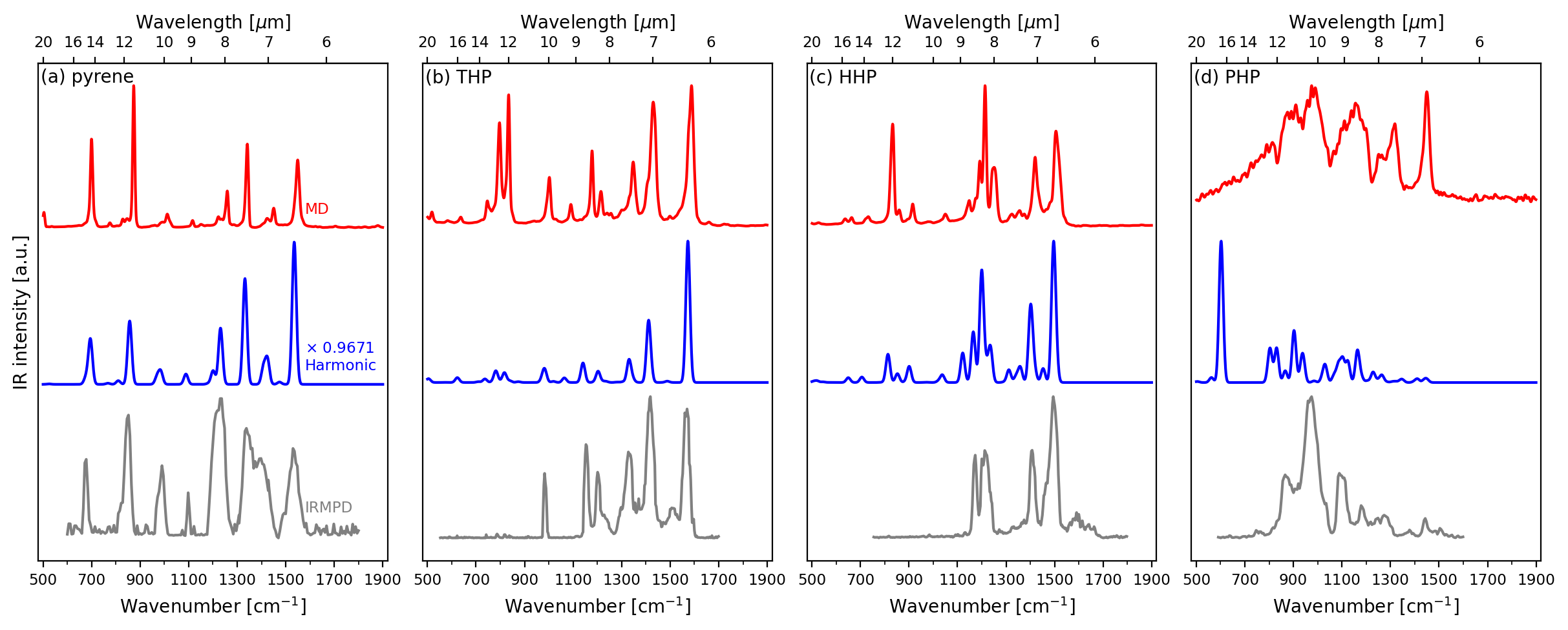}
	\caption{Comparisons between theoretical and experimental IR spectra of (a) pyrene; (b) THP; (c) HHP and (d) PHP.
	Red lines are obtained from MLMD simulations, while blue lines are from harmonic calculations using the MLIP and a empirical scaling factor of 0.9671 is applied on harmonic frequencies.
	Experimental IR spectra are shown in gray lines.
	Each spectrum has been normalized to its maximum peak intensity.
	}
    \label{fig:anharmonic-ir}
    \end{center}
\end{figure*}

The applicability of the harmonic approximation in the IR spectra of pyrene, THP, HHP and PHP cations, are investigated by comparing theoretical IR spectra (via the harmonic approximation and MD simulations) with experimentally measured IR spectra.
Such a comparison is shown in \cref{fig:anharmonic-ir}, in which all IR spectra have been normalized to the maximum IR absorption intensity.
All theoretical IR spectra are obtained in this work and are computed with the MLIP.
The experimental IR spectrum of the pyrene cation is taken from Panchagnula et al. \cite{panchagnula2020}, and that of THP cation is taken from Simonsen et al \cite{simonsen2024}.
In this study, the experimental IR spectra of the HHP and PHP cations are measured with the IRMPD spectroscopy \cite{IRMPD_review,IRMPD_tutorial_review}.

In the case of pyrene cation, the experimental IR spectrum agrees well with the harmonic IR spectrum when an empirical scaling factor of 0.9671 \cite{kesharwani2015} is applied to harmonic frequencies. 
The MD-based spectrum resembles band positions and relative intensities of the scaled harmonic spectrum.
For the partially superhydrogenated pyrene cations (THP, HHP), the experimental band positions are well reproduced by both the scaled harmonic spectrum and the MD-based spectrum.
In terms of relative band intensities, MD simulations tend to predict higher intensities for C-H out-of-plane bending modes below 900 cm$^{-1}$, compared with those obtained within the harmonic approximation.
When the pyrene cation gets fully superhydrogenated (PHP), the harmonic approximation fails in reproducing the overall band profile obtained by the experiment.
For example, the harmonic spectrum produces the most intense band around 600 cm$^{-1}$ while the experimental spectrum has extremely low intensities for bands below 800 cm$^{-1}$.
The MD-based spectrum provides a more favorable match with the IRMPD spectrum for the PHP cation.
Similar behavior is also observed in the IR spectrum of the indeyl cation (C$_9$H$_7$$^+$), in which MD-generated spectrum gives a much better match with the experiment than the harmonic scaled spectrum. \cite{wenzel2025}

\subsection{Role of temperature}
\begin{figure}
	\begin{center}
	\includegraphics[width=\columnwidth]{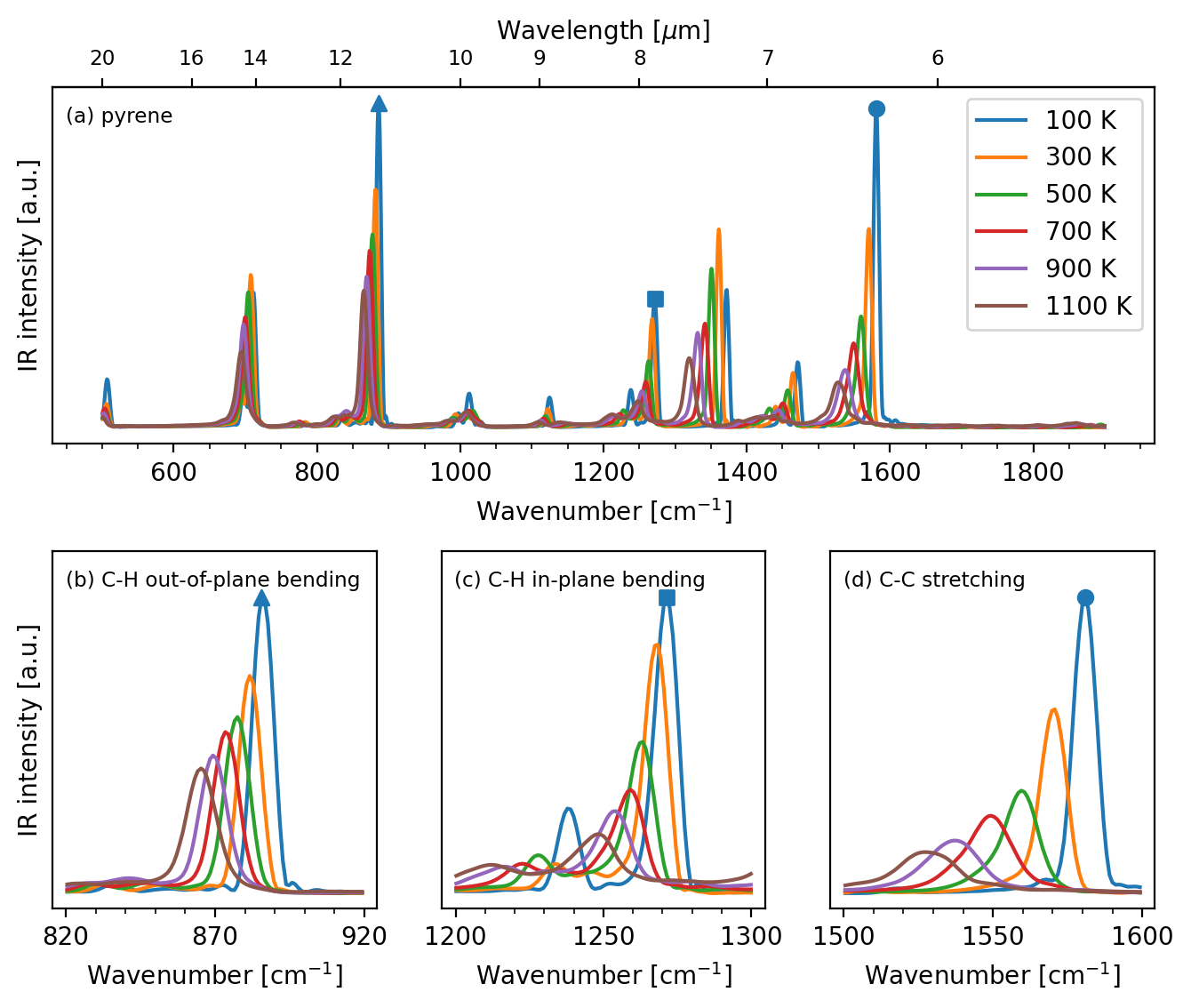}
	\caption{Temperature-dependent IR spectra of the pyrene cation from MLMD simulations. 
	(a) Frequency ranges from 500 to 1900 cm$^{-1}$ 
	(b) the C-H out-of-plane bending mode in the frequency region 820 - 920 cm$^{-1}$
	(c) the C-H in-plane bending mode in the frequency region 1200 - 1300 cm$^{-1}$
	(d) the C-C stretching mode in the frequency region 1500 - 1600 cm$^{-1}$
	}
    \label{fig:pyrene-ir}
    \end{center}
\end{figure}

MLMD simulations are performed at low and high temperatures (100, 300, 500, 700, 900 and 1100 K), in order to study the role of temperature in IR spectra of superhydrogenated pyrene cations.
The variations in band positions and widths are studied in details to quantify the role of temperature.
The pyrene cation is used as the first example and its temperature-dependent IR spectra is shown in \cref{fig:pyrene-ir}.
A general trend is that, as temperature increases, vibrational modes shift to lower frequency and get broader.
Meanwhile, IR intensity becomes weaker.

\begin{figure}
	\begin{center}
	\includegraphics[width=\columnwidth]{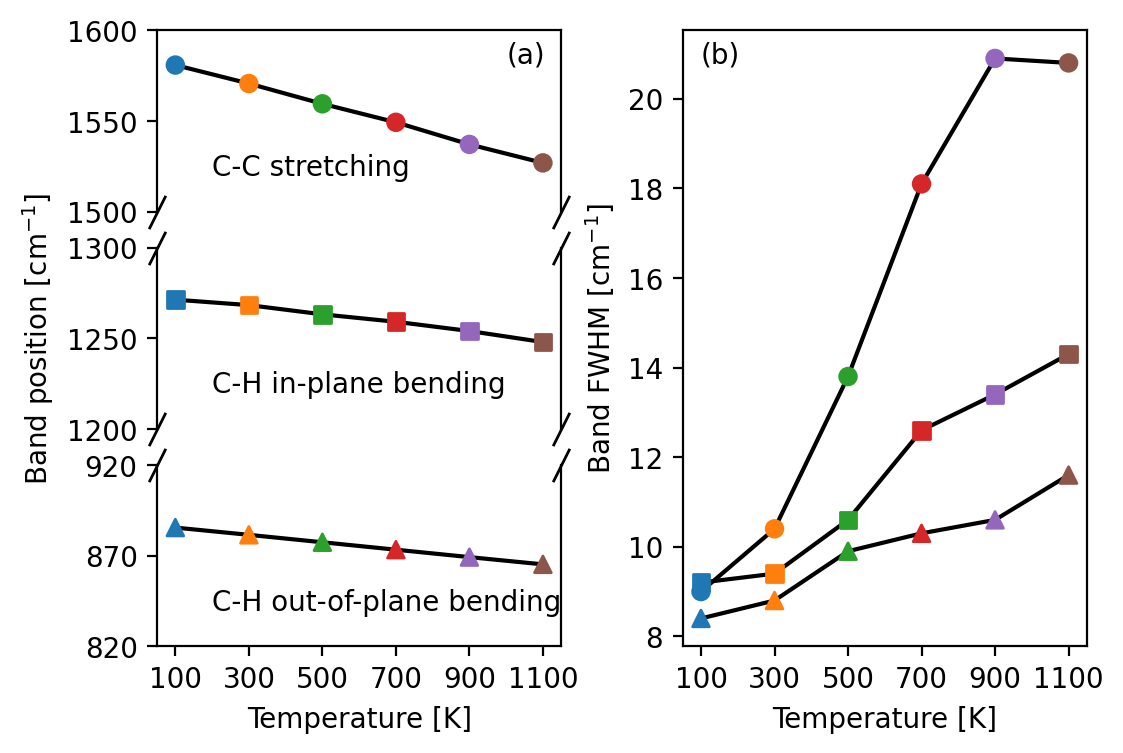}
	\caption{Quantitative analysis of the role of temperature for three modes in the pyrene cation. 
	(a) band position in cm$^{-1}$
	(b) band full width at half maximum (FWHM) in cm$^{-1}$
	}
    \label{fig:pyrene-pos-width}
    \end{center}
\end{figure}

\cref{fig:pyrene-pos-width} presents a quantitative analysis of the thermal evolution of band positions and band widths (full width at half maximum, FWHM) 
of three typical PAH vibrational modes (C-H out-of-plane bending, C-H in-plane bending and C-C stretching).
Three bands all show linear shifts to lower frequencies at higher temperatures, with their slopes normally defined as the anharmonicity coefficients $\chi ^ \prime$ \cite{joblin1995}.
Among those three bands, the C-C stretching has the highest band position and also the highest $\chi ^ \prime$ of $-$0.054 cm$^{-1}$K$^{-1}$ in the temperature range 100$-$1000 K,
followed by the C-H in-plane bending with $-$0.023 cm$^{-1}$K$^{-1}$ and the C-H out-of-plane bending with $-$0.020 cm$^{-1}$K$^{-1}$.
Those red-shifts are comparable with that in the C-H stretching region of neutral pyrene (approximately $-$0.020 cm$^{-1}$K$^{-1}$). \cite{joblin1995}
The red-shifts are normally interpreted as being due to vibrational level spacing becoming smaller at higher internal energies in an anharmonic potential well. 
All three bands get broader as temperature increases.
Similarly, the C-C stretching has the largest broadening, compared with the C-H in-plane and out-of-plane bendings.

\begin{figure*}
	\begin{center}
	\includegraphics[width=\textwidth]{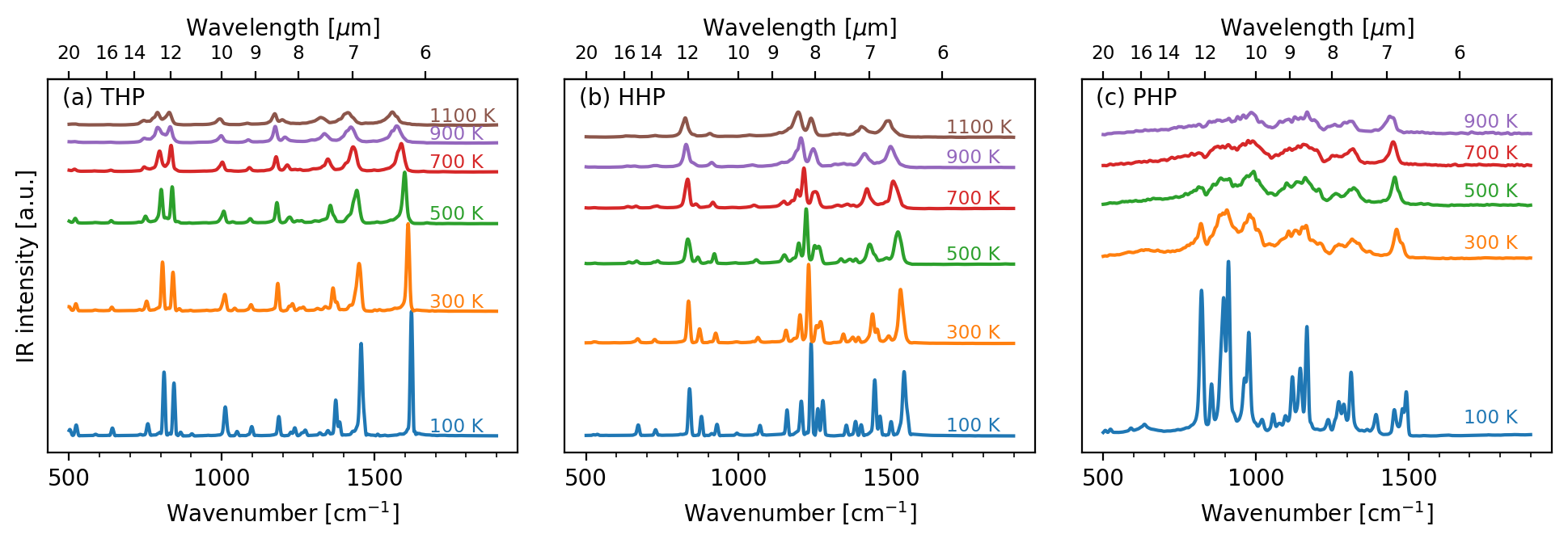}
	\caption{Temperature-dependent IR spectra of (a) THP; (b) HHP and (c) PHP from MLMD simulations.
	}
    \label{fig:hpyrenes-ir}
    \end{center}
\end{figure*}

\cref{fig:hpyrenes-ir} shows IR spectral evolution with temperatures for THP, HHP and PHP.
THP in \cref{fig:hpyrenes-ir} (a) and HHP in \cref{fig:hpyrenes-ir} (b) have similar temperature dependence as that of the pyrene cation, 
namely linear red-shifts and broadenings at higher temperatures.
More band mergings are observed in THP and HHP, possibly due to a larger vibrational degrees of freedom when the extra H atoms are added on the pyrene skeleton.
When the pyrene cation becomes fully superhydrogenated into PHP, there is a dramatic change more than a linear evolution in the band profile from 100 K to 300 K, as shown in \cref{fig:hpyrenes-ir} (c).
The change includes the transformation from a discrete number of sharp bands with high IR intensities to a few broad and unresolved features with low IR intensities.
The spectral difference may reflect the variation in the sampling in the configuration space.
Some dynamics at 300 K cannot be captured by the MLMD simulation at 100 K.
This may explain the failure of harmonic approximation in PHP, since small displacements near the equilibrium configuration are not enough to describe PHP's vibrations.
It it necessary to run MLMD with the temperature higher than or equal to 300 K in order to obtain a reliable prediction of PHP's IR spectrum.
As the temperature goes beyond 300 K, there is no dramatic evolution in the band profile.
At 1100 K, PHP gets fragmented during the MLMD simulation.
This is possibly due to PHP's thermal instability at 1100 K.
Alternatively, our training data set has an insufficient sampling of configurations near 1100 K
The detailed reason requires further investigations, which are beyond the scope of this work.

\section{Astrophysical implications}
Recently, several nitrogen-substituted pyrenes have been detected in the dense cloud TMC-1 by radio telescopes \cite{wenzel2024, wenzel2024a}.
It is estimated that up to 0.1 \% of carbon in TMC-1 could be locked up in pyrene \cite{wenzel2024}.
In terms of infrared observations, a deeper understanding of vibrational features in pyrene-based PAHs could be helpful for interpreting JWST spectra.
A JWST study of the Orion bar concludes that the band profile of 6.2 $\mu$m AIBs (normally assigned as aromatic C-C stretching modes in cationic PAHs) is controlled by anharmonicity \cite{chown2024}.
Our calculations of the pyrene cation have shown that, the vibrational peak of the aromatic C-C stretching mode, has shifted linearly from 1580.9 cm$^{-1}$ (6.32 $\mu$m) at 100 K to 1526.9 cm$^{-1}$ (6.55 $\mu$m) at 1100 K.
Its FWHM has increased from 9.0 cm$^{-1}$ at 100 K to 20.8 cm$^{-1}$ at 1100 K, both of which are narrower than that of the 6.2 $\mu$m AIBs, e.g. 34.8 cm$^{-1}$ in the H II region \cite{chown2024}.
For pyrene cation which only has aromatic carbon, it is possible to fit empirical anharmonicity factors to describe linear band shifts and broadenings as temperature increases.
This becomes difficult for partially superhydrogenated pyrene cations (THP and THP), since their bands become more blurred and less isolated when the temperature increases.
In the upper limit of superhydrogenated pyrene cation, the PHP in which all carbon atoms are aliphatic, a single empirical factor cannot capture the complicated blending of bands at increased temperatures.
Temperature-dependent IR spectra of PHP have to be computed on a case-by-case basis.

\section{Conclusion}
We have performed MD simulations with the MLIP in order to investigate the applicability of the harmonic approximation and the role of temperature in the IR spectra of the pyrene cation and its superhydrogenated derivatives.
The harmonic approximation with a empirical scaling factor is able to explain the experimental band profile of the pyrene cation and its partially superhydrogenated derivatives (HHP and THP).
However, it fails in the case of PHP which only has aliphatic carbon atoms but no aromatic carbon atoms.
Anharmonic treatment via the MD-based approach is necessary to obtain a reliable IR spectrum of the PHP cation.
As the temperature increases, the band positions of the pyrene cation shift to lower frequencies and the absorption features get broader.
The temperature dependence of band positions and widths is almost linear for the pyrene cation.
The linear trend becomes less evident as the degree of superhydrogenation increases.
The PHP cation shows a dramatic change in the band profile from 100 K to 300 K, which is not linear anymore.

\begin{acknowledgments}
This work has been supported by the Danish National Research Foundation through the Center of Excellence “InterCat” (Grant agreement no.: DNRF150), the European Union (EU) and Horizon 2020 funding award under the Marie Skłodowska-Curie action to the EUROPAH consortium, grant number 722346.
We acknowledge support from VILLUM FONDEN (Investigator grant, Project No. 16562).
Z.T. acknowledges support from the Research Start-up Fund Project of Hainan University (No. XJ2400011241).
\end{acknowledgments}

\section{Data Availability}
The MLIP model developed in this work and spectral data in this article are publicly available in a Zenodo repository (\href{https://zenodo.org/records/15228144}{https://zenodo.org/records/15228144}).
Other data that support the findings of this work are available within the article.

\section*{References}
\bibliography{ref}

\end{document}